\documentclass[aps,prd,10pt,superscriptaddress,tightline,nofootinbib,numerical,floatfix,twocolumn]{revtex4-2}
\usepackage{stackengine}
\usepackage{bm}
\usepackage{braket}
\usepackage{color}
\usepackage{lipsum}
\usepackage{epsfig}
\usepackage{float}
\usepackage[T1]{fontenc}
\usepackage{graphicx}
\usepackage[latin9]{inputenc}
\usepackage{mathtools}
\usepackage{xcolor}
\usepackage{cleveref}
\usepackage{amssymb}
\usepackage{amsthm}
\usepackage{amsmath}
\usepackage[font={small}]{caption}

\begin{document}

\title{Entanglement based tomography to probe new macroscopic forces }

\newcommand{\affone}{Centre for Quantum Computation and Communication Technology, School of Mathematics and Physics, University of Queensland, Brisbane, Queensland 4072, Australia}
\newcommand{\afftwo}{Department of Physics and Astronomy, University College London, Gower Street, WC1E 6BT London, United Kingdom.}
\newcommand{\affthree}{Van Swinderen Institute, University of Groningen, 9747 AG Groningen, The Netherlands.}
	
\author{Peter. F. Barker}
\affiliation{\afftwo}

\author{Sougato Bose}
\affiliation{\afftwo}

\author{Ryan J. Marshman}
\affiliation{\affone}
	
\author{Anupam Mazumdar}
\affiliation{\affthree}
	
\date{\today}

\begin{abstract}
Quantum entanglement provides a novel way to test short distance physics in the non-relativistic regime. We will provide a protocol to {\it potentially} test new physics by bringing two charged massive particle interferometers adjacent to each other. Being charged, the two superpositions will be entangled via electromagnetic interactions mediated by the photons, including the Coulomb and the Casimir-Polder potential. We will bring a method of {\it entanglement based tomography} to seek time evolution of very small entanglement phases to probe new physical effects mediated by {\it hitherto unknown macroscopic force} which might be responsible for entangling the two charged superpositions modelled by the Yukawa type potential. We will be able to constrain the Yukawa couplings $\alpha \geq 10^{-35}$ for $r\geq 10^{-6}$m for new physics occurring in the electromagnetic sector, and in the gravitational potential  $\alpha_g \geq 10^{-8}$ for $r \geq 10^{-6}$m. Furthermore, our protocol can also constrain the axion like particle mass and coupling, which is complimentary to the existing experimental bounds.
\end{abstract}
\maketitle

One critical observation which parts from the classical world is the notion of quantum entanglement~\cite{Schrodinger,Horodecki}. The latter provides the evidence of quantum correlation which a classical world cannot replicate. In particular, it is known that a classical interaction cannot entangle the two quantum systems (if they were not entangled to begin with)~\cite{LOCC}. Since all the known Standard Model (SM) interactions and mediators are quantum in nature, {\it entanglement} formation becomes inevitable.  The SM interactions are fairly well constrained by the collider~\cite{PDG}, non-collider~\cite{Jaeckel:2010ni}, the electron-dipole-moment (EDM)~\cite{Chupp:2017rkp} experiments. However, there are consistent effort in constraining weakly coupled bosons such as axion like particle, majorons,  or light dark matter searches~\cite{Essig:2013lka,Graham:2015ouw}. Given these advancements it is now paramount to seek a complimentary avenue to test new physics in the infrared (IR) and in a non-relativistic (NR) limit.

The aim of this paper will be to provide a  simple quantum-information lead {\it entanglement}  protocol to confirm the SM interactions and to probe new physics in the particle physics and gravitational sectors. Very light and weakly coupled bosons could couple to the SM degrees of freedom, such as axion like particle, Kaluza-Klein modes from the extra dimensions, hidden sector photon~\cite{Moody,Essig:2013lka,Jaeckel:2010ni,Graham:2015ouw,Klimchitskaya:2021lak}. 
Axions with varied mass range are also expected in string theory~\cite{Svrcek:2006yi}, which provide a significant motivation to look for these light bosons experimentally. In the NR limit such corrections will generically yield Yukawa modifications to the potential mediated by these light bosons either at the tree or at the loop level~\cite{Moody}. Similarly, the quantum nature of the graviton within general relativity (GR) and beyond GR also tend to modify the gravitational potential which can be tested via {\it entanglement witness}~\cite{Bose:2017nin,Marshman:2019sne,Marletto,Bose:2022uxe}. 

In this paper, we will probe the nature of beyond the SM physics and the gravitational sector  by witnessing the quantum entanglement between the two charged massive particle interferometers. Specifically, we will assume the interferometers induce a spatial superposition of size $\Delta x$. For details about explicit schemes of the sorts of intereferometers considered, see Refs.~\cite{Bose:2017nin,Marshman:2021spm,Margalit:2020qcy}.  

To model this, we take an electronic spin in a host crystal, with a spin-spatial state coupling allowing us to measure the entanglement. We will assume a nitrogen-vacancy (NV) centre spin in a diamond nano/micro-crystal, although qualitatively our results and analysis will hold for similar materials. We will further assume that the nano-crystal can be electrically charged.  If we bring another such quantum superposition of a charged nano-crystal and keep them apart by a distance $d$, then the two nano-crystals will be entangled via the exchange of a virtual photon (Coulomb), via the vacuum induced dipole-dipole type interactions mediated by the two photon exchange giving rise to the Casimir-Polder (CP) potential~\cite{CP,Holstein}, and any additional particle-particle interactions which may occur. Since, at sufficiently short distances both the contributions will dominate the electromagnetic (EM) interactions, we study how to disentangle or mitigate the EM induced entanglement to probe new macroscopic forces in the EM sector and in the gravitational sector

In this paper, we will show that there exists parameter regions, where we can cancel the entanglement phase due to the Coulomb and the CP potential, and probe the configuration where the net entanglement phase due to the EM induced interactions vanishes (when $\Delta \phi_{em}= 2n\pi$, for $n=0,1,2\cdots$). We will then show how by studying the entanglement entropy in the bipatrite system, namely the von-Neumann entropy, we can probe for unknown force of nature in the IR. 
To detect a new force, we consider that the resulting particle-particle interactions produce different quantum correlations, which are captured by the time evolution of the {\it entanglement entropy}, i.e. {\it entanglement based tomography}. In general, we will describe this new force by a Yukawa potential without going into the details of model building. 

\begin{figure}
	\centering
	\includegraphics[width=0.8\linewidth]{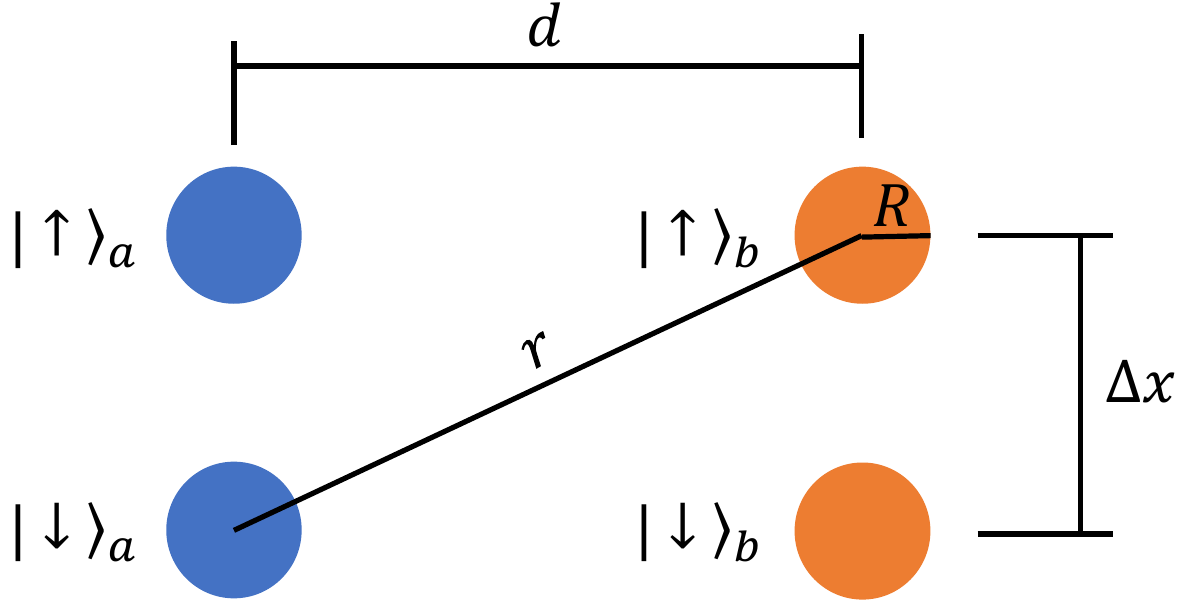}
	\caption{Configuration where the two spatial superpositions with the splitting $\Delta x$ are kept parallel to each other separated by a distance $r,~d$. The two spin states have been shown, the radius of the  spherical nano-crystal is $R$.  }
	\label{fig:system configuration}
\end{figure}
\begin{figure}
 \includegraphics[width=1.0\linewidth]{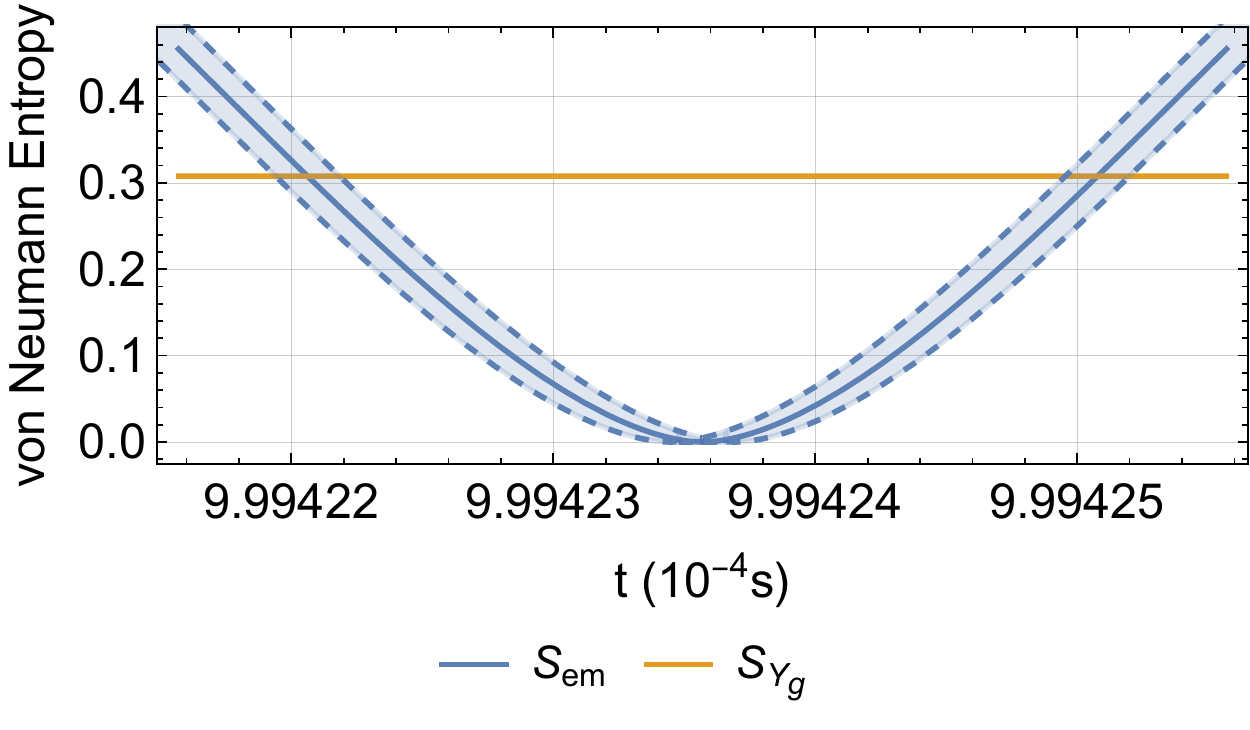}		
 \caption{von Neumann entropy with respect to time for all the interactions. The shaded region shows the uncertainty in the experimental parameters where the EM induced entanglement is used as a baseline. Here  we have taken $R=10^{-4} {\rm m}, \alpha_g=0.01, \lambda=10^{-3}$.} 
 \label{fig:entropy}
\end{figure}

Let us consider a setup where we have two charged nano-crystals, with an embedded, addressable spin, we will assume that they have the same mass $m$ with a radius $R=\left({3m}/{4\pi\rho}\right)^{1/3}$ where $\rho=3.5\times10^{3}$ kgm$^{-3}$ is the density of the nano-crystals, similar to that of the diamond, see Fig.~(\ref{fig:system configuration}).  Each mass, labelled $a$ and $b$, will be assumed to have a net charge $(q_{a}e,~q_{b}e)$ where $q_i\in\mathbb{Z}$ and $e$ is the charge of an electron. The interactions are expected to be in the IR and are thus limited to EM, gravitational and the unknown interactions mediated by beyond SM quanta. Since the nano-crystals have diamagnetic properties, typical for a diamond, there will be a CP potential on top of the Coulomb interaction. These will be the dominant EM interactions the crystals will face at relatively short distances, but still in the IR.  The known potentials are
\begin{eqnarray}
U_{cp}(x) &=& -\frac{23\hbar c}{4\pi} \frac{\epsilon - 1}{\epsilon + 2}\frac{R^6}{\left(x-2R\right)^7} \\
U_{c}(x) &=& \frac{e^2}{4\pi\varepsilon_0}\frac{q_aq_b}{x}
\end{eqnarray}
where $\epsilon$ is the diamagnetic susceptibility.
We consider the masses to be placed in a spatial superposition of size $\Delta x$ in the parallel arrangement as shown in Fig.(\ref{fig:system configuration}). For this configuration of masses, the joint quantum state of the spins $\left|\Psi(0)\right\rangle = \frac{1}{2}\left(\left|\uparrow,\uparrow\right\rangle + \left|\downarrow,\downarrow\right\rangle + \left|\uparrow,\downarrow\right\rangle + \left|\downarrow,\uparrow\right\rangle\right)$ will evolve to~\cite{Bose:2017nin,Marshman:2019sne}~\footnote{The analogous system of quantum spatial superposition but with the neutral masses have been considered in Ref.~\cite{Bose:2017nin,Marshman:2019sne}. The details of the entanglement phase evolution can be found in Refs.~\cite{Bose:2017nin,Marshman:2019sne,vandeKamp:2020rqh}. In these papers, the quantum nature of the graviton (spin-2 and spin-0 components of the quantum nature of graviton was responsible for entangling the two superpositions, see for a quantum field theory description~\cite{Marshman:2019sne,Bose:2022uxe}. Here, instead, we consider the quantum nature of the photon to entangle the two charged nano-crystals. Even the vacuum induced dipole-dipole interaction which gives rise to CP potential is being mediated by the virtual exchange of the photons~\cite{Holstein}.}:
\begin{equation}
    \left|\Psi(t)\right\rangle =\frac{1}{2}\left[\left|\uparrow,\uparrow\right\rangle + \left|\downarrow,\downarrow\right\rangle + e^{i\Delta\phi(d,r)}\left(\left|\uparrow,\downarrow\right\rangle + \left|\downarrow,\uparrow\right\rangle\right)\right] \nonumber
\end{equation}
where the phase $\phi$ is determined by the interaction considered: $\phi_{i}(x) = ({t}/{\hbar})U_{i}(x)$, see~\cite{Marshman:2021spm,Bose:2022uxe}, and is a function of the particle-particle distance $d$ between the $\left|\uparrow\right\rangle_a$ and $\left|\uparrow\right\rangle_b$ states (which is the same as that for the joint state $\left|\downarrow,\downarrow\right\rangle$) and $\Delta\phi(d,r)=\phi(r)-\phi(d)$ where $r=\sqrt{d^2+\Delta x^2}$ is the particle-particle distance for the joint states $\left|\uparrow,\downarrow\right\rangle$ (and equivalently $\left|\downarrow,\uparrow\right\rangle$). As such, there is only one important phase $\Delta\phi(d,\Delta x)$, which dictates whether or not the masses are entangled, which is why this arrangement is preferred. Further to these potentials, we will consider as an example of the Yukawa potential where we will constrain  $(\alpha, \lambda) $, see~\cite{Moody,Klimchitskaya:2021lak}.
\begin{equation}
	U_{Y}(x)= {\alpha e^{-{x}/{\lambda}}}/{x}\,,
\end{equation}
$\alpha$ dictates the interaction strength of the new physics and $\lambda$ determines the effective range of the interaction, or related to the Compton wavelength $\lambda \sim 1/m_\ast$, where $m_\ast$ denotes the particle mass which interacts with the EM photons.  We will now discuss how, by carefully selecting the experimental parameters, we might hope to detect new close range forces and the modifications to these by eliminating the competing effects of the Coulomb and the CP interactions.
To determine the parameter space in which the Coulomb and the CP interactions cancel, we will consider the entanglement phase due to each:
\begin{eqnarray}
\Delta\phi_{c} &=& \frac{t}{\hbar} \frac{e^2}{4\pi\varepsilon_0}\left(\frac{q_aq_b}{r} - \frac{q_aq_b}{d}\right)\,,\nonumber\\
\Delta\phi_{cp} &= &-\frac{t}{\hbar} \frac{23\hbar c}{4\pi} \frac{\epsilon - 1}{\epsilon + 2}
\left(\frac{R^6}{\left(r-2R\right)^7} - \frac{R^6}{\left(d-2R\right)^7}\right).
\end{eqnarray}
%
Thus, we require $\Delta\phi_{c}(d, r)=-\Delta\phi_{cp}(d, r)$. 
%
To consider a general situation, we use the parametrisation; $\Delta x = \beta d$, $r=\sqrt{1+\beta} d$. This constrains
\begin{equation}
d = \left(\frac{\sqrt{1+\beta}-\left(1+\beta\right)^{-3}}{\sqrt{1+\beta}-1}\frac{23\hbar c\varepsilon_0}{e^2} \frac{\epsilon - 1}{\epsilon + 2} \frac{1}{q_aq_b}\right)^{1/6}R.
\end{equation}
These results suggest that there are no solutions when $\textrm{sign}(q_a)\ne\textrm{sign}(q_b)$ as this leads to a complex valued $d$. The ability to detect a close range interaction within this regime will then be determined by the parameter stability, and the certainty from one run to the next. There is however a significant issue of the acceleration induced by the CP and the Coulomb interactions.

 The two accelerations fail to cancel one another out, and leave a significant residual acceleration, which is only made larger by increasing the charges of the masses ($q_a$ and $q_b$). This poses a strong limiting factor without some further mitigation strategy, which is what we will now present.




Now, we will present the analysis of the experimental configuration where we optimise the separation between the two superpositions involved, such that the CP and the Coulomb forces cancel. So that we are no longer optimising over both the average distance $d$ and $\Delta x$. The two EM forces are given by; $F_{cp}(x) = -({161\hbar c}/{4\pi}) ({\epsilon - 1}/{\epsilon + 2}){R^6}/{\left(x-2R\right)^8}$, and 
$F_{c}(x) = ({e^2}/{4\pi\varepsilon_0})({q_aq_b}/{x^2}) $.
%
We can eliminate the net EM forces at a single distance $d$, which gives $F_{cp}(d) = -F_{c}(d)$. Setting $d=nR$ gives:
\begin{align}
		\frac{n^2}{\left(n-2\right)^8}  =&  \frac{4\pi}{161\hbar c} \frac{\epsilon + 2}{\epsilon - 1}\frac{q_aq_be^2}{4\pi\varepsilon_0} \,,\label{eq:force cancellation distance}
\end{align}
 which gives $d\sim 6R, ~{\rm for}~~ q_a=q_b=1$, 
	$d\sim 4R,~{\rm for}~~ q_a=q_b=10$,~ and  
   $ d\sim 3R,~{\rm for}~~ q_a=q_b=100$.

\begin{figure}
	\centering
	\includegraphics[width=1.0\linewidth]{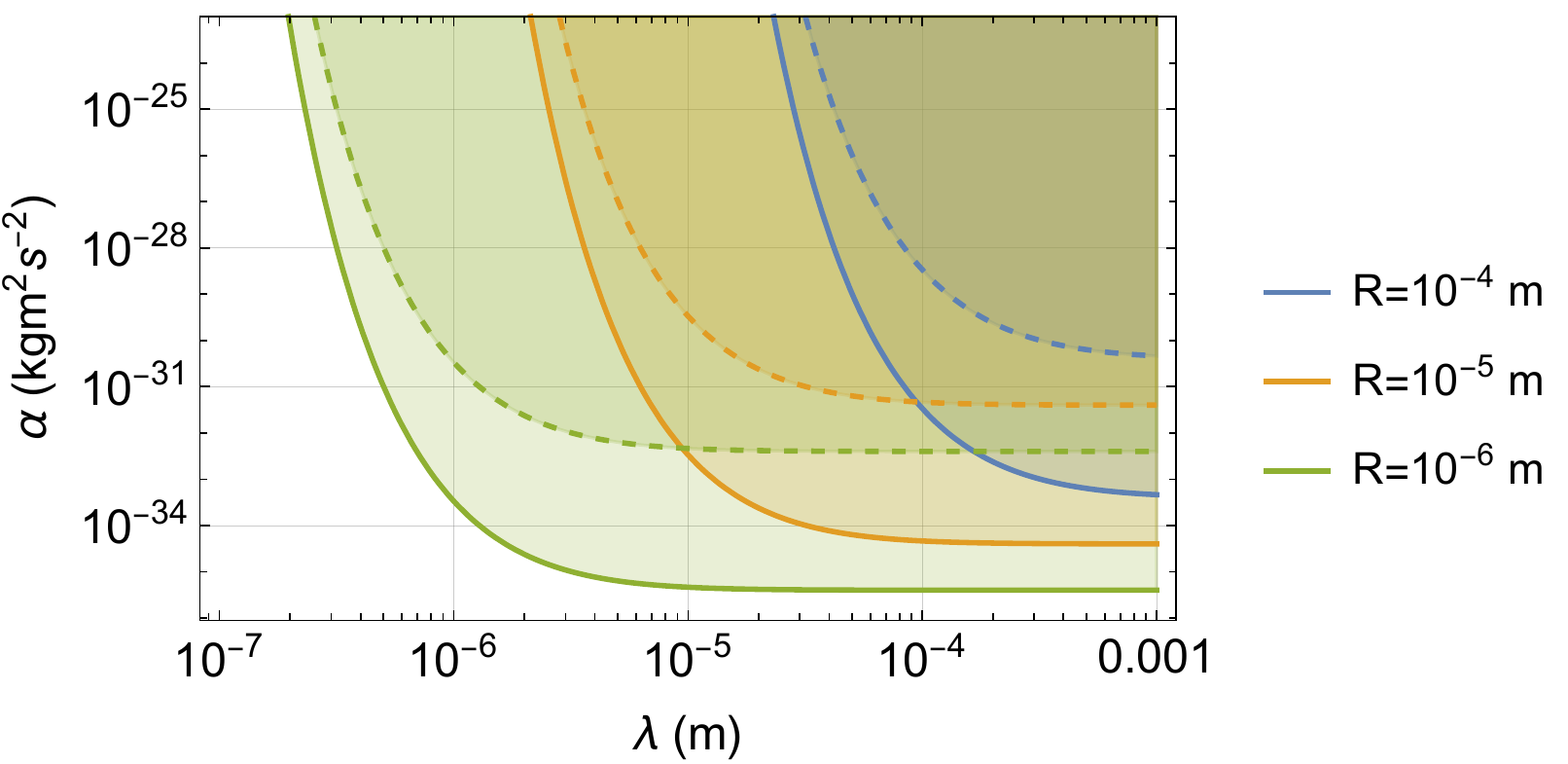}
	\caption{$\alpha-\lambda$ plot showing the region in which the Yukawa potential dominates over the Casimir and Coulomb interactions for different values of $R$. The solid and dashed lines correspond to a total interaction time $t=10^{-3}, t=10^{-6}$ s, respectively. \label{fig:alpha-lambda plot}}
\end{figure}

\begin{figure}
	\includegraphics[width=1.0\linewidth]{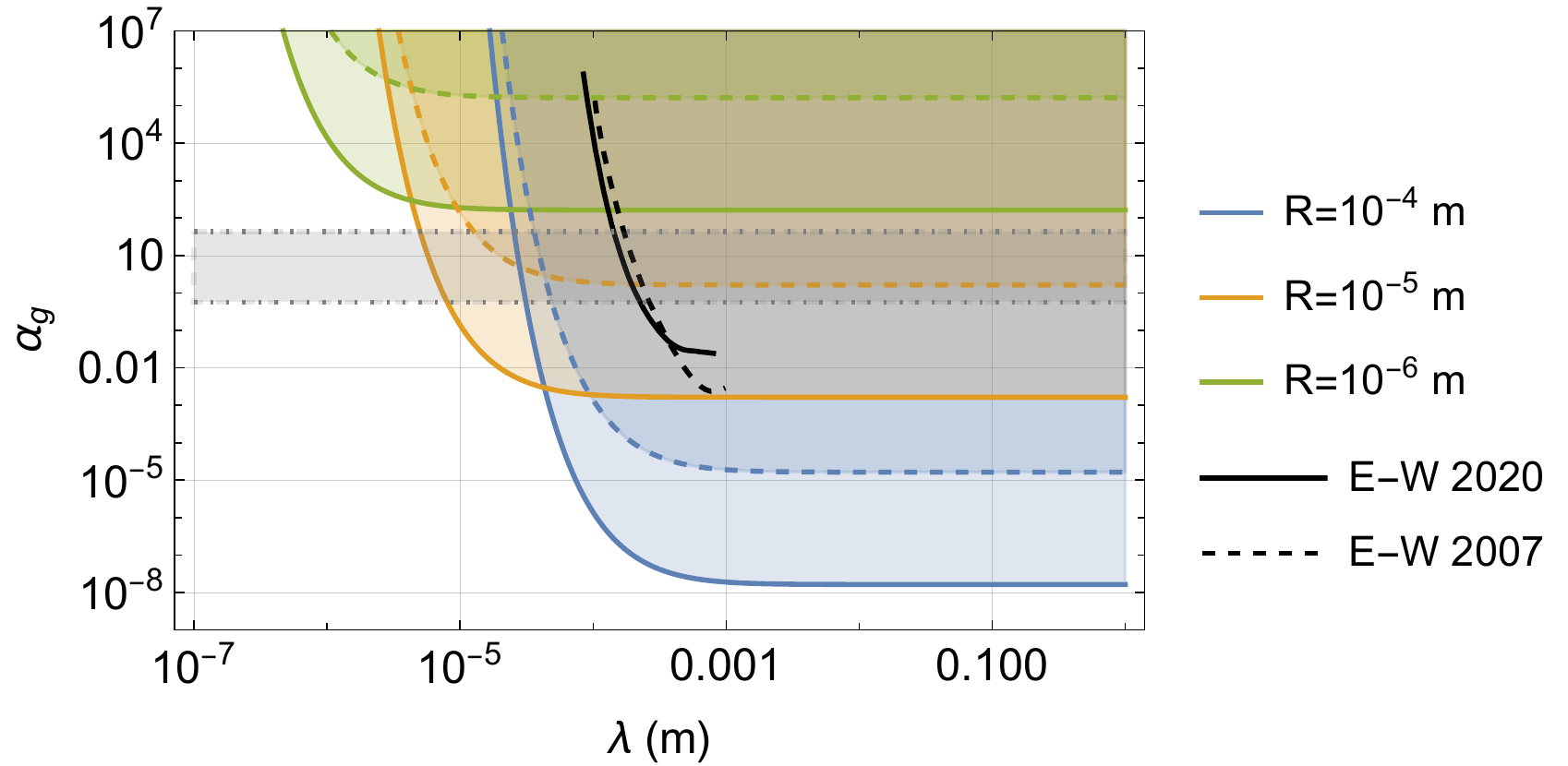}
	\caption{$\alpha_g-\lambda$ plot showing the region in which the Yukawa modified Newtonian potential dominates over the Casimir and the Coulomb interactions for different values of $R$. The solid and dashed lines correspond to the total interaction time $t=10^{-3}, t=10^{-6}$ s, respectively. The two black lines (solid and dashed) correspond to the experimental results from the years 2007, 2020, see~\cite{Lee:2020zjt}. The horizontal grey area shows the region in which $n$-compact extra dimensions may be observable where $\alpha_g=2n$. The typical range for $\alpha_g \leq 20$ from string theory compactification on a Calabi Yau manifold, see~\cite{Kehagias:1999my}. }
	\label{fig:alphaG-lambda plot}
\end{figure}

This limits the charges to be, $q\lesssim {\cal O}(10)$. To allow a significant separation between the two nano-crystals, we will need to move into the regime where the forces only {\it approximately} cancel. To do this, we would set the minimum distance between the masses, $d$, to be such that the force between them cancels. 
The superposition size $\Delta x$ can then be increased until the force between the distant states (each at a distance of $r$ from one another) becomes significant. We will again use $d=nR$ to represent the exact distance for which the forces cancel, and write $r=nR+r'$. The residual force is then
\begin{equation}
	F_{\textrm{net}}=\left(\frac{161\hbar c}{4\pi} \frac{\epsilon - 1}{\epsilon + 2}\frac{8}{\left(n-2\right)^9} - \frac{e^2q_aq_b}{4\pi\varepsilon_0}\frac{2}{n^3}\right)\frac{r'}{R^3}
\end{equation}
Therefore, for example, when we take $q_a=q_b=1$, and $n=6$, we obtain $F_{\textrm{net}}\sim10^{-29}{r'}/{R^3}$. This suggests that, depending on the tolerable force, even $r'\leq R$ could be possible. In this case, the superposition size is $\Delta x =((nR + r')^2 -n^2R^2)^{1/2}$. For the purpose of illustration and in remainder of the paper, and specifically for Figs.(\ref{fig:entropy},\ref{fig:alpha-lambda plot},\ref{fig:alphaG-lambda plot}), we will take these optimised parameters,
\begin{equation}
d=6R,~r=7R,~\Delta x\approx 3.6R, ~q_1=q_2=1\,.
\end{equation}
 We consider $R\in\left\{10^{-4}\textrm{ m},10^{-5}\textrm{ m},10^{-6}\textrm{ m}\right\}$, equivalently $m\in\left\{10^{-8}\textrm{ kg},10^{-11}\textrm{ kg},10^{-14}\textrm{ kg}\right\}$. 	
It is, however, known that for a given mass $m$, the graviton vacuum gets displaced and this 
corresponds to the number of excited gravitons around the Minkowski vacuum which can be estimated by the Bekenstein's entropy~\cite{Bekenstein}: $N_g=S_{BEK}\sim (m/M_p)^2$~\cite{Bose:2021ekc}. For $m\sim 10^{-8}$kg, it is actually $N_g\sim 1$, and for lighter mass $N_g \ll 1$. It appears
that $N_g \sim {\cal O}(1)$ might dictate this subtle distinction between one graviton excitation with many graviton excitations in the vacuum. Here we have kept all our masses below the Planck mass, i.e. $m\leq 10^{-8}$kg.

We will now consider a novel strategy to mitigate the EM induced phase through the tomography, i.e. experimental timing. Specifically, given the total EM-induced entanglement phase is $\Delta\phi_{em}=\Delta\phi_{c}+\Delta\phi_{cp}$, and provided the experimental parameters ($m$, $t$, $d$ and $\Delta x$) are tuned sufficiently well, then ensuring
\begin{equation}
	\Delta\phi_{em}=\Delta\phi_{c}+\Delta\phi_{cp}=  2\pi n \label{eq:no em phase condition}
\end{equation}
where $n\in\mathbb{Z}$ will remove any EM signal, leaving only the discovery potential for the other macroscopic forces beyond the SM.

This can be achieved by choosing an appropriate interaction time. If we are constraining the Yukawa potential, then by using our set-up we can utilise the force cancelling region to enable a relative stable particle-particle interaction. That is, little or no deflection of the masses due to their EM interactions throughout the interferometry process. This can be seen in Fig.~(\ref{fig:entropy}), where the entanglement entropy induced by the EM interaction between the two sub-systems vanishes, while the other interactions do not vanish at a specific point. Indeed, we have chosen a small snap-shot of time for the purpose of illustration, we can indeed take a larger time evolution. Just as an example, we have shown here the entanglement entropy for the  gravitational induced Yukawa potential between the two systems do not vanish at that specific time. The shaded region in the EM induced entanglement denotes the uncertainties in the Coulomb and CP potential due to uncertainties in the experimental parameters, as discussed below.

The EM induced entanglement phases per unit time are given by:
\begin{eqnarray}
	\frac{\Delta\phi_{c}}{t} &=& \frac{1}{\hbar} \frac{e^2}{4\pi\varepsilon_0}\left(\frac{q_aq_b}{r} - \frac{q_aq_b}{d}\right) 
	\sim ({5\times10^{4}}/{R}) \textrm{rad s}^{-1} \nonumber\\
	\frac{\Delta\phi_{CP}}{t} &=&\frac{1}{\hbar} \frac{23\hbar c}{4\pi} \frac{\epsilon - 1}{\epsilon + 2}\left(\frac{R^6}{\left(r-2R\right)^7} - \frac{R^6}{\left(d-2R\right)^7}\right) \nonumber\\
	&\sim&-({2\times10^{4}}/{R}) \textrm{rad s}^{-1}. \label{eq:em phases}
\end{eqnarray}
For any interaction time $t\gtrsim10^{-4}R$~sec, the condition set out in Eq.(\ref{eq:no em phase condition}) can be readily made true. We will consider the merit of two different scenarios of new physics. 

\noindent
 $\bullet$ {\it Departure from the Coulomb and the CP potential:} \\
 We will consider the acceptable levels of uncertainties in the experimental parameters. To determine, approximately, we will consider the uncertainties by taylor expanding arond their target values which suggests the Coulomb interaction will be 
\begin{eqnarray}
\Delta\phi_{c}(d, r) + \delta\left(\Delta\phi_{c}(d, r)\right)=\nonumber \\
	\Delta\phi_{c}(d, r)(1+{\delta t}/{t} + {\delta d}/{d} + {\delta r}/{r})
	 \nonumber \\
	 \equiv  \Delta\phi_{c}(d, r)(1+3\tilde{\delta}),
\end{eqnarray}
where we have assumed every parameter has an equal relative error $\tilde{\delta}$ for simplicity. 
Similarly for the CP interaction, we obtain
\begin{equation}
	\Delta\phi_{cp}(d, r)+\delta\Delta\phi_{cp}(d, r)\approx \Delta\phi_{cp}(d, r)(1+15\tilde{\delta}).
\end{equation}
 
To ensure that the entanglement phases induced by the EM interactions does not remove that sourced by the Yukawa, $\delta\Delta\phi_{em} \ll \Delta\phi_{Y}$ and $\delta\Delta\phi_{em} \ll 1$, where $\Delta \phi_Y$ is the entanglement phase induced by the Yukawa term. Thus we will require $\left|\delta\Delta\phi_{em}\right|< 1$, which implies $\tilde{\delta}\left|\left(3\Delta\phi_{c} + 15\Delta\phi_{cp} \right)\right|< 1$, or using Equation~\ref{eq:em phases}: $\tilde{\delta} \le10^{-6} {R}/{t}$. Similarly, we must ensure that the gravitational interaction between the masses do not induce a coherence destroying noise. Given the gravitational entanglement phase is given by: $\Delta\phi_{g}={Gm^2t}\left(r^{-1}-d^{-1}\right)/\hbar$, to keep the variance below the unit phase will require $\delta\Delta\phi_{g}\approx\frac{40\pi^2\rho^2GR^5t}{189 \hbar}\tilde{\delta}< 1$. Substituting known values and assuming diamond is used gives then gives the bound $\tilde{\delta}\lesssim 5 \times 10^{-32}t^{-1}R^{-5}$. If we thus set $\tilde{\delta}\approx \textrm{min}\left\{10^{-6} {R}/{t},10^{-32}t^{-1}R^{-5}\right\}$, then the Yukawa potential will be the dominant entangling signal, provided $\left|\Delta\Phi_y\right|\ge1$. We then have a detectable Yukawa signal provided
\begin{equation}
	\frac{\alpha t}{42\hbar R}\left(6e^{-\frac{7R}{\lambda}}-7e^{-\frac{6R}{\lambda}}\right)\geq 1\,.
	 \label{eq:alpha lambda sensitivity}
\end{equation}
The Yukawa induced correction to the EM interaction is shown in Fig.~\ref{fig:alpha-lambda plot}. Its worth noting that this corresponds to the actual parameter uncertainties of:
$	\delta d= \textrm{min}\left\{10^{-6}\times({R^2}/{t}){\rm sec~m^{-1}},~10^{-32}\times t^{-1}R^{-4}{\rm sec~m^{5}}\right\},$ and $ \delta t= \textrm{min}\left\{10^{-6}\times R\textrm{ sec~m}^{-1},~10^{-32}\times R^{-5}\textrm{ sec~m}^{5}\right\}$.
This suggests that for a nano-crystal with a radius $R\sim 10^{-6}$m we will need the precision in time for the measurement to be around $10^{-12}$~sec. This may pose a challenge but the recent advancements of keeping track of the frequency ratio measurements at 18-digit accuracy may be the way to track the time evolution of the entanglement~\cite{BACON}.

\begin{figure}
	\includegraphics[width=1.0\linewidth]{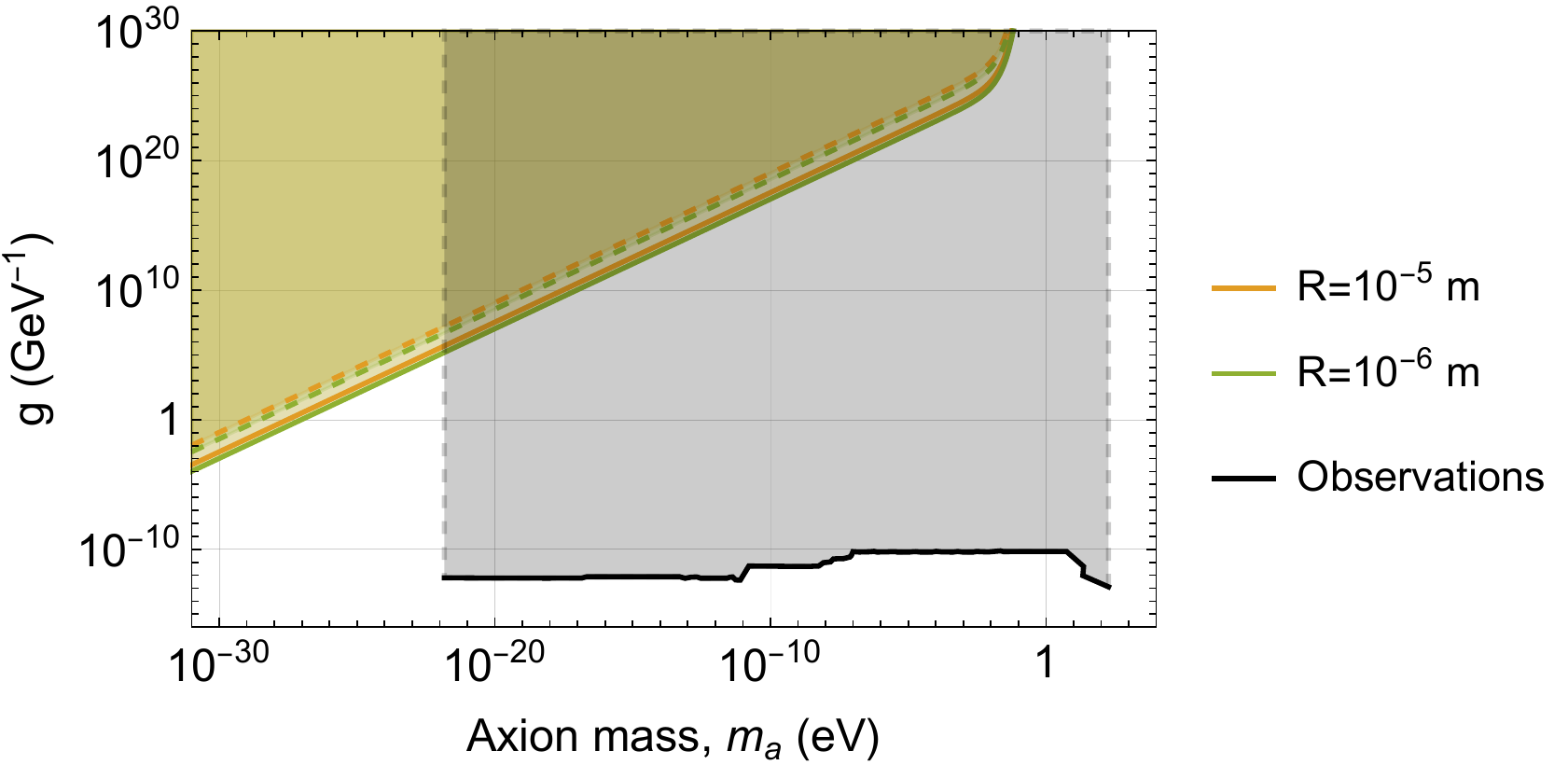}
	\caption{Plot showing the region in which the Axion modified Coulomb interaction is detectable using this entanglement based detection for the two configurations. The \emph{Observations} shown by the grey shaded region correspond to the combined results of many separate observations, such as CAST+SUMICO, cosmology, supernovae-1987A, electron-positron collider (LEP), OSCAR and PVLAS collaborations, compiled from
	the Refs.~\cite{Jaeckel:2010ni, Essig:2013lka, Alekhin, Arik, Capolupo:2019peg, Villalba-Chavez2018}. \label{fig:axion plot}}
\end{figure}

$\bullet$ {\it Axion like particle detection: }
We can also consider how our protocol may be sensitive to the axion modified Coulomb potential. Following previous analyses~\cite{Villalba-Chavez2018}, assuming that the experiment involves distances larger than the Compton wavelength of the axion mass,  the axion modified Coulomb potential is approximately given by: $U_{ac}(x)=U_c(x)\left(1+\frac{g^2 \lambda^2 \hbar^2}{\pi^2c^2 x^4}e^{-x/\lambda}\right)$ where $g$ is the coupling strength and $\lambda=\frac{\hbar}{m_a c}$ is the Compton wavelength of the axion of mass $m_a$. The entanglement phase detectability requires $\Delta\phi_{ac} - \delta\left(\Delta\phi_{ac}\right) >  \Delta\phi_{c} + \delta\left(\Delta\phi_{c}\right)$, which leads to $ \Delta\phi_{a} > 2\delta\left(\Delta\phi_{c}\right)$. The latter condition gives the bound for a detectable $g$ as
\begin{equation}
	g\gtrsim\sqrt{ 6\frac{\pi^2c^2}{\lambda^2\hbar^2}\left(\frac{1}{r}-\frac{1}{d}\right)\left(\frac{e^{-r/\lambda}}{r^5}-\frac{e^{-d/\lambda}}{d^5}\right)^{-1}\tilde{\delta}}.
\end{equation}
We can see that the current observational bounds arising from \cite{Jaeckel:2010ni, Essig:2013lka, Alekhin, Arik, Capolupo:2019peg, Villalba-Chavez2018} cover a large range of parameter space in $g,~m_a$. We see that our current protocol will be sensitive to constrain masses below $m_a\leq 10^{-21}$~eV in future.

As such, taking the experimental set-up discussed above with $R=10$ $\mu$m suggests a the detectable coupling strength will be
\begin{equation}
	g\gtrsim \frac{10^{35}}{\lambda}\left(2 e^{-\frac{6}{10^{-5}\lambda}}-e^{-\frac{7}{10^{-5}\lambda}}\right)^{-1/2}\textrm{ kg}^{-1}
\end{equation}
as shown in Figure \ref{fig:axion plot}.

$\bullet$ {\it Departure from the Newtonian potential:} Following the above discussion, we can evaluate the entanglement phase as sourced by the quantum origin of the gravitational interaction~\cite{Bose:2017nin,Marshman:2021spm,Bose:2022uxe} or as we will do here, a modification to the Newtonian potential. We can parameterise the departure from Newton's law by the Yukawa potential~\cite{Adelberger:2003zx}:
\begin{equation}\label{mod-newt}
	U_{Y_g}=\frac{Gm^2}{r}\left(1+\alpha_g e^{-{r}/{\lambda}}\right)
\end{equation}
In this case, a similar analysis as above, see Eq.(\ref{eq:alpha lambda sensitivity}) translates to a sensitivity in the effective gravitational coupling $\alpha_g$ given by Fig.(\ref{fig:alphaG-lambda plot}). Note that the sensitivity to probe $\alpha_g$ in the vicinity of $10^{-5}$~m is far improved than what we have achieved in the torsion-balance experiments $\alpha_g\sim 10^{-3}$ in the ${\mu m}$ range~\cite{Lee:2020zjt}. Fig.(\ref{fig:alphaG-lambda plot}) shows the experimental constraints from 2000 and 2007 of the same group, see~\cite{Lee:2020zjt}.

Large extra spatial dimensions~\cite{Antoniadis:1997zg,Arkani-Hamed:1998jmv} also modify the gravitational potential very similar to the Yukawa modification, Eq.(\ref{mod-newt}), see~\cite{Kehagias:1999my}. The extra dimensions are compactified, which depends on the geometry of the compactification. However, different compactification will give rise to different $\alpha_g$ in Eq.(\ref{mod-newt}). If $n$ extra dimensions are compactified on a n-torus, then $\alpha_g =2n$, when on n-spehere $\alpha_g=n+1$,  and in the case of string theory compactification on a Calabi-Yau manifold it should be  $\alpha_g\leq 20$~\cite{Kehagias:1999my}. The corresponding $\alpha_g-\lambda$ plot is shown in Figure \ref{fig:alphaG-lambda plot} which suggests that our setup with entanglement tomography is capable of detecting modifications to the Newtonian potential for $\lambda\gtrsim 10^{-6}$ m, and eventually can probe certain aspects of string theory or the presence of large extra dimensions~\footnote{Similarly, we can also constrain the hidden sector photon, for which the modified Coulomb potential can be derived   very similar to the modification to the gravitational potential~\cite{Kroff:2020zhp}, $V(r)=[e^2/(4\pi \epsilon_0 r)][1+\alpha e^{-m_{\gamma'}r}]$. Here $m_{\gamma'}$ is the mass of the hidden photon and $\sqrt{\alpha} $ designates the coupling between the hidden sector photon and the Standard Model photon, i.e. $(\sqrt{\alpha}/2)X_{\mu\nu}F^{\mu\nu}$.  $X_{\mu\nu}$ denotes the hidden sector photon and $F_{\mu\nu}$ denotes the Standard Model photon, and $\mu\nu =0,1,2,3$. In this scenario we can constrain $m_{\gamma'}\leq 10^{-17}$~eV for $\alpha \geq 10^{-4}$. The current constraints on the hidden sector photon mostly arises from astrophysics, and they can constrain $m_{\gamma'}\sim {\cal O}(10^{-10}-10^{-15})$eV, see~\cite{Kroff:2020zhp}. }

To summarise, all the examples capture the importance of entanglement based tomography as a protocol to test new physics - whether it is in the EM  sector or  in the gravitational sector. Indeed, there are many challenges from creating charged superposition~\cite{Wineland} in a levitated ion trap, see~\cite{Barker,Penny:2021gwj,Wei,Romero,Delrod,Hsu}, to preparing and cooling the ground state  of the nano-crystal~\cite{Delic}. There will be constraints on the decoherence~\cite{Basi}, but we expect the decoherence rates will be very similar to the case of the neutral crystal, as far as the blackbody and the atmospheric interactions are concerned for a micron size crystal with a mass of 
$\sim10^{-17}$Kg~\cite{vandeKamp:2020rqh,Tilly:2021qef}. However, there are possible ways of mitigating EM induced decoherence by creating Faraday cage, and also the gravity gradient and relative acceleration noise~\cite{Toros:2020dbf}. 

To conclude, the entanglement based tomography could potentially be a powerful protocol to detect new physics in the IR. One can apply this technique to probe modifications to GR including local~\cite{Addazi:2021xuf} and non-local modifications of classical and quantum gravity~\cite{Biswas:2011ar,Tomboulis,Edholm:2016hbt}. In physics beyond the SM, we will be able to probe certain model dependent parameters case by case~\cite{Moody,Graham:2015ouw,Capolupo:2019peg}. The last  reference considered axion induced entanglement, however, not in the context of EM background effects arising from Coulomb and CP-induced entanglement.

In a model independent fashion the entanglement based tomography can constrain the Yukawa parameters  $\alpha \geq 10^{-35}$ for $r\geq 10^{-6}$m for the new forces, and a similar modification in the gravitational potential would yield constraining $\alpha_g \geq 10^{-8}$ for $r \geq 10^{-6}$m, nearly five-orders of magnitude better than the current bounds arising from the classical torsion-balanced experiment~\cite{Lee:2020zjt}. In this way we can also place future constraints on the large extra compact dimensions. Furthermore, in the  case of axion like particle detection, we are sensitive to even smaller axion masses (smaller than the existing constraints from various astrophysical and cosmological observations), but not for small couplings, see Fig.~(\ref{fig:axion plot}). Nevertheless, entanglement tomography in conjunction with other data set will be extremely helpful to probe axion masses below $10^{-21}$eV, thereby probing both Yukawa type correction as well as more sophisticated potentials arising due to axion-like particles.

{\it Acknowledgements}:
	PFB. acknowledges funding from the EPSRC Grant No. EP/N031105/1 and the H2020-EU.1.2.1 TEQ project Grant agreement ID: 766900. SB acknowledges the EPSRC grant EP/S000267/1, AM acknowledges (NWO) grant number 680-91-119 and RJM was supported by the Australian Research Council (ARC) under the Centre of Excellence for Quantum Computation and Communication Technology (CE170100012).


\end{document}